\documentclass[10pt]{ismd08}
\usepackage{graphicx}
\def\pppbarp{$pp/{\bar p}p$}

\def\p{$p$} 
\begin{document}
\title{ISMD08 \\  Photoproduction total cross-sections
at very high energies and the Froissart bound}
\author{Y. N. Srivastava$^1$\protect\footnote{\\ speaker},
A. Achilli$^1$, R. Godbole$^2$, A. Grau$^3$, G. Pancheri$^4$}
\institute{$^1$INFN and Physics Department, University of Perugia, I-06123 Perugia, Italy
\\
$^2$ Centre for High Energy Physics, Indian Institute of Science,Bangalore, 560012, India
\\
$^3$Departamento de F\'\i sica Te\'orica y del Cosmos,
Universidad de Granada, 18071 Granada, Spain
\\
$^4$INFN Frascati National Laboratories, I-00044 Frascati, Italy
}
\maketitle
\begin{abstract}
A previously successful model for purely hadronic total cross-sections, based on QCD minijets and soft-gluon resummation, is here applied to the   total photoproduction  cross section. We  find  that 
our model in the $\gamma p$ case predicts a rise with energy stronger than in the \pppbarp \  case. 
\end{abstract}
\section{Introduction}
In this note, we shall  describe (and apply to data)  a model for  the total cross-section \cite{ourmodel,prd72},
based on the ansatz that infrared gluons provide the saturation mechanism in the rise of all  total cross-sections (thus obeying the Froissart bound), with  the rise calculated through the increasing number of hard collisions between  low-x, but perturbative gluons. These  collisions produce  low $p_t$ partons which hadronize in so called  mini-jets. We assume that for  $p_t\ge 1\div 2 \ GeV$ the parton-parton cross-section can still be calculated perturbatively and set a minimum $p_t$ cut-off, $p_{tmin}$ in the jet cross-section calculation. To make connection with actual phenomenological inputs,  the mini-jet cross-sections are calculated \cite{Pramana,plb659} using DGLAP evolved parton densities: for the proton we have used   GRV \cite{GRV},   MRST \cite{mrst} and CTEQ \cite{cteq}, for the photon GRS \cite{grs99ph} and CJKL \cite{cjkl}.
In our model we use only LO densities, as part of the NLO effects are described by soft gluon resummation  and the use of NLO would result in some double counting. Similarly, we have opted for tree level parton-parton cross-sections and one loop $\alpha_s$. As the c.m. energy increases, with fixed $p_{min}$, these mini-jet cross-sections increase and their contribution to the total cross-section becomes larger than any observed cross-section, violating unitarity. This has resulted in discarding  the mini-jet model.  Embedding the mini-jet cross-section in the eikonal representation, restores unitarity, but requires modelling of the matter distribution in the colliding particles via an impact parameter distribution. Convolution of the electromagnetic  form factors is frequently used and more fundamental attempts exist in the framework of Reggeon calculus and perturbative QCD. Our model focuses on very soft gluons as the source of a dynamical description of the impact factor and its energy dependence. Thus the name Bloch Nordsieck (BN) underlies the infrared region and its resummation. 
We shall briefly present this model, show its results for purely proton processes,  and then apply it to photoproduction processes.

\section{The Bloch-Nordsieck Model (BN)}
Our BN model exhibits fractal behaviours for quantities such as (i) the energy rise of the mini-jet cross-sections for which  $\sigma_{jet} \approx s^ \delta $ with $\delta \approx 0.3$ and (ii) the very low momentum single gluon emission probability which we propose to be  proportional to $k_t^{-p-1}$ with $0<p<1$, for gluons of transverse momentum $k_t$  .  
 
   This model, which was initially developed  for purely hadronic total cross-section, incorporates QCD inputs such as parton-parton cross-sections, realistic parton densities, actual kinematics, and soft gluon resummation.
We write, for a general process, 
\begin{equation}
\sigma^{AB}_{tot} =2 \int d^2 {\vec b}[1-e^{-n(b,s)/2}]
\label{sigtot}
\end{equation}
with the imaginary part of the eikonal related to the average number of inelastic collision $n(b,s)$. We isolate all hard perturbatively calculated collisions into 
\begin{equation}
n_{hard}(b,s)=A(b,s)\sigma_{jet}(p_{tmin},s)
\end{equation}
and phenomenologically determine the remaining collisions which we call $n_{soft}(b,s)$. At present our model is unable to make an {\it ab initio} calculation of this quantity, and we use a QCD inspired modelling, described in \cite{prd72}.

The impact parameter distribution is obtained from the Fourier transform of the soft gluon transverse momentum, resummed, distribution, namely
\begin{equation}
 A(b,s) =
{\cal N} \int d^2 {\bf K}_{\perp} {{d^2P({\bf K}_\perp)}\over{d^2 {\bf K}_\perp}}
 e^{-i{\bf K}_\perp\cdot {\bf b}} \nonumber \\
 = {{e^{-h( b,q_{max})}}\over
 {\int d^2{\bf b} e^{-h(b,q_{max})} }}\equiv A(b,q_{max}(s))
 \label{Eq:abn}
 \end{equation}
with
 \begin{equation}
h( b,q_{max}(s))  =\int d^3{\bar n}_g({\vec k})
[1-e^{
i{\bf k}_\perp\cdot {\bf b}
}
]=
\frac{16}{3}\int_0^{q_{max}(s) }
{{dk_t}\over{k_t}}
 {{ \alpha_s(k_t^2) }\over{\pi}}   \left(\log{{2q_{\max}(s)}\over{k_t}}\right)\left[1-J_0(k_tb)\right]
\label{hdb}
\end{equation}
In the above equation, we need to extend the integral to zero momentum gluons, which supply the saturation effect of resummation. One needs then an ansatz for the  single soft gluon distribution in Eq. \ref{hdb}, namely for $\alpha_s(k_t^2)$ as $k_t \rightarrow 0$. Our model for this behaviour is inspired by the Richardson potential, but  in order to have a finite result for the integral in   Eq. \ref{hdb} we use
\begin{equation}
\alpha_s={{12 \pi}\over{33-2N_f}} {{p}
\over{\ln[1+p({{k_t}\over{\Lambda}})^{2p}]}}.
\label{alphas}
\end{equation}
This expression gives the asymptotic freedom value for large $k_t$ , and is singular (but integrable) at $k_t=0$ (for $p<1$). The  closer \p  \ is to 1,  the more the minijet cross-sections will be quenched at any given energy.

The energy dependence of the impact function $A(b,s)$ is introduced through the upper limit of integration in Eq. \ref{hdb}. As amply discussed in ref. \cite{ourmodel}, the function $q_{max}$ is obtained through an averaging over the parton densities
 \vspace*{0.5cm}
\begin{equation} 
  q_{max}(s)= \sqrt {\frac{s} {2}}\
\frac{{\sum\limits_{i,j} {\int {\frac{{dx_1 }} {{x_1 }}\int
{\frac{{dx_2 }} {{x_2 }} \int_{z_{min}}^1 {dz f_i (x_1) f_j (x_2)
\sqrt {x_1 x_2 } (1 - z)} } } } }} {{\sum\limits_{i,j} {\int
{\frac{{dx_1 }} {{x_1 }}\int {\frac{{dx_2}} {{x_2}}
\int_{z_{min}}^1 {dz} f_i (x_1)f_j (x_2) } } } }},
\label{qmax}
\end{equation}
Notice that in our model, the impact parameter distribution depends on the energy and the process under consideration  through the parameter $q_{\max } (s)$, which is evaluated using the given parton densities. 

\begin{figure}[htb]  
\centering
\includegraphics*[scale=0.30]{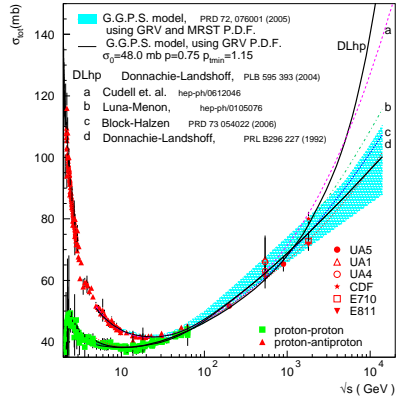}
\caption{Data and models for proton-proton and proton-antiproton total cross-section from ref.
\protect\cite{plb659}.
}
\label{lhcfig}
\end{figure}

The BN model thus described has been applied to proton-proton scattering, obtaining a  total cross-section for LHC  to be $\sigma(\sqrt{s}=14\ TeV)= (100 \pm 12) \  mb$, where the error reflects various uncertainties as in the choice of densities, minimum parton $p_t $ cut-off and the IR behaviour of the soft gluon coupling. 
Our results for proton-proton and proton-antiproton scattering are shown in Fig. \ref{lhcfig}  with labelling and references defined as in \cite{plb659}.
\section{Photon processes and the total $\gamma p$ cross-section at high energies }
Application to photons requires the probability that a photon behaves like a hadron. One possibility is to use Vector Meson Dominance (VMD)
in the eikonal representation, as in  \cite{sarcevic,halzen},
\begin{equation}
\sigma^{\gamma p}_{tot} =P_{\gamma \rightarrow hadron}  \sigma^{\gamma_{had} p}_{tot}=2 P_{had}\int d^2 {\vec b}[1-e^{-n(b,s)/2}]
\label{sigtotgp}
\end{equation}
with $P_{\gamma \rightarrow hadron}=1/240$. 
 As for the proton case, the average number of inelastic collisions, $n(b,s)$,  is split between hard collisions calculable as QCD minijets, and a soft part. Hence, the average number of collisions is written as
\begin{equation}
n(b,s)=n_{soft}(b,s) +n_{hard}(b,s) =  {{2}\over{3}}  n_{soft}^{p p}(b,s)+ A(b,s) \sigma_{jet}(s)/P_{had}
\end{equation}
with  $n_{hard}$ including all outgoing parton processes with $p_t>p_{tmin}$. The jet cross-sections are calculated using actual  photon densities, which themselves give the probability of finding a given quark or gluon in a photon, and thus  $P_{had}$ needs to be  canceled out in $n_{hard}$.
For  the soft part, a good description is obtained with $n_{soft}^{p p}(b,s)$ being  the same as in the $pp$ case\cite{plb659}.
The impact function $A(b, s)$ again supplies saturation and is calculated using photon and proton densities in Eq. \ref{qmax}. Once this energy parameter has been calculated, $A(b,s)$ is fully determined. More fundamental attempts to obtain the impact function for photons can be found in ref. \cite{bartels}.

\begin{figure}[htbp]  
\centering
\includegraphics[scale=0.30]{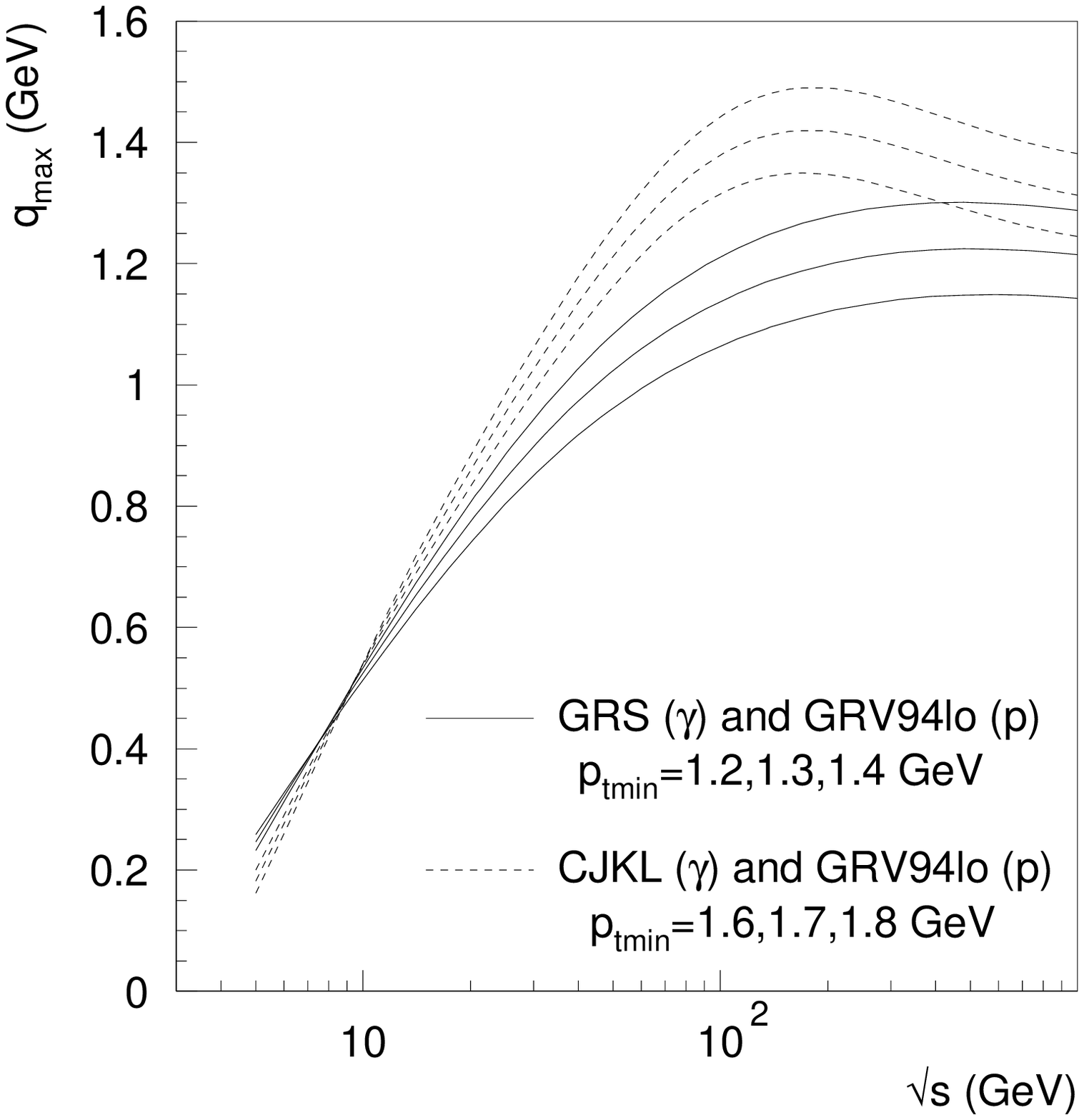}
\hspace{.5cm}
\includegraphics[scale=0.3]{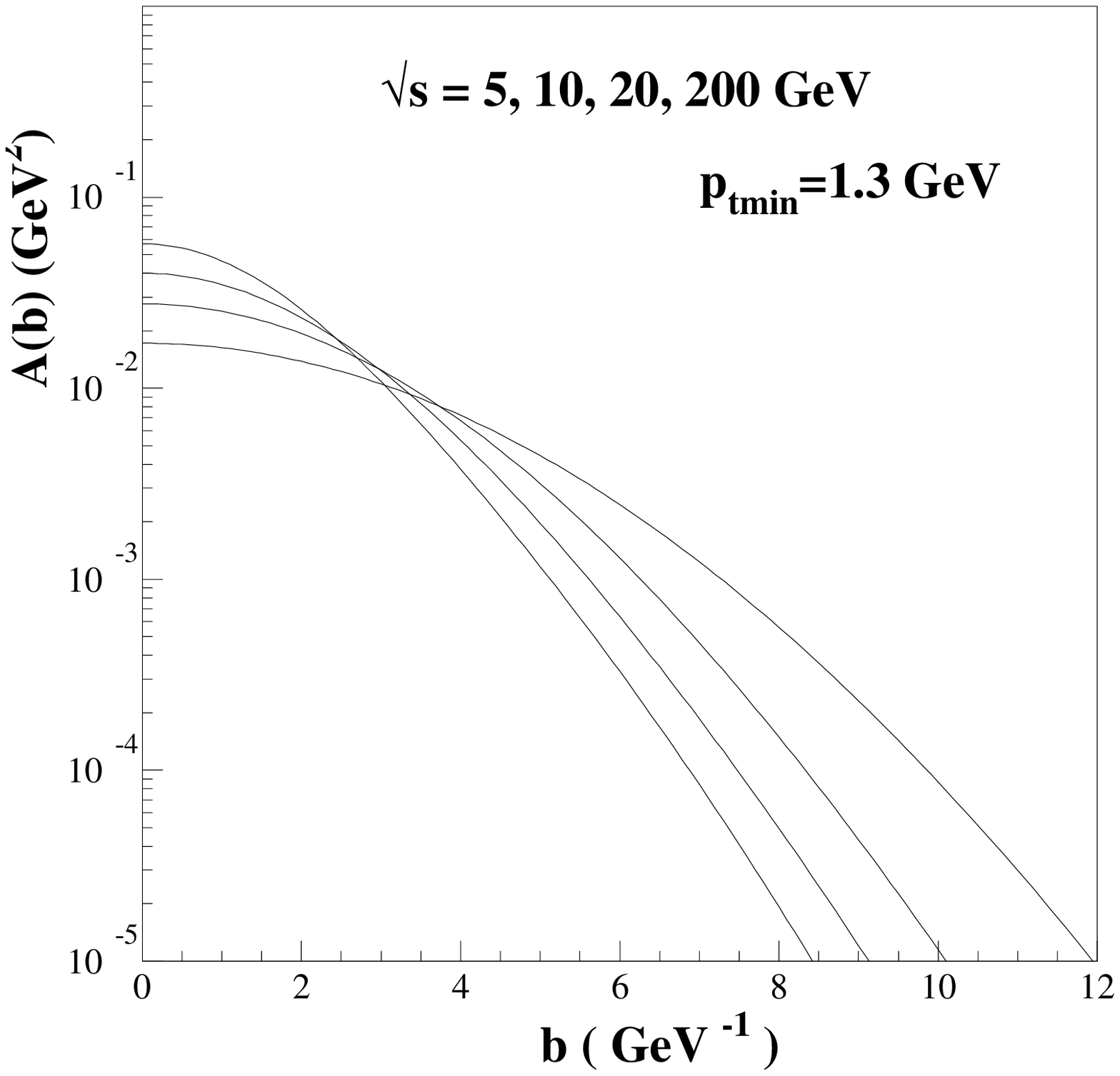}
\hspace {.1cm}
\caption{Left: average maximum tramsverse momentum allowed to single soft gluon emission in $\gamma p$ scattering. 
Right: Impact parameter distribution for $\gamma p$ scattering at various c.m. energies}
\label{qmaxadb}
\end{figure}

For $\gamma p $  we show in Fig. \ref{qmaxadb} both the saturation parameter $q_{max}$ plotted as a function of the $\gamma p $ c.m. energy, as well as the resulting impact parameter function at four representative energies. Unlike other models, based on the convolution of the form factors of the colliding particles, the impact function in the BN model is  energy dependent, with a shape shrinking with energy.
Both the  mini-jet cross-section and the impact function (the latter through $q_{max}$) for $\gamma p$ depend on the set of Parton Density Functions (PDFs). We choose GRV94 for the proton, GRS and CJKL for the photon, input them into the eikonal and compare the results with HERA data \cite{h1,zeus}, including  a set of ZEUS BPC data extrapolated from $\gamma^* p$  to $Q^2_\gamma=0$ \cite{haidt}. These results for the BN model are shown in  Table \ref{tablesigma} for different parameters sets chosen so that all of HERA data are included in a band defined by the last two columns in table \ref{tablesigma}.

In ref. \cite{gampnew}, we have observed that at very high $\gamma p$ energies, our results indicate a faster rise than is the case for proton inspired models. For energies beyond HERA, $q_{\max}(s)$, computed through the photon densities, no longer increases (unlike the proton case) thus blocking saturation earlier than for protons. As a result, since the mini-jet cross-sections keep on increasing, the photon cross-sections, past present accelerator energies,  would grow faster than the purely hadronic ones.
We have noted  \cite{gampnew} that this prediction from our model finds independent support in 
the fit by Block and Halzen \cite{bh} which  gives results close to ours in the very high energy region.

\begin{footnotesize}
\begin{table*}[hbtp!]
\caption{Values (in $m b$) for total cross-section for $\gamma p$ 
scattering evaluated in the c.m. energy of colliding particles, 
for different parameter sets}
\label{tablesigma}
\begin{center}
\begin{tabular}{||c||c|c|c|c||} \hline \hline
$\sqrt{s} $
&EMM with Form 
&$BN_\gamma$ model 
&$BN_\gamma$ model 
&$BN_{\gamma}$ \\
 GeV & Factors                       &GRS, p=0.75       &CJKL, p=0.8      &GRS, p=0.75 \\ 
          &$p_{tmin}=1.5$ GeV & $p_{tmin}=1.2$ &$ p_{tmin}=1.8$ & $p_{tmin}=1.15$\\
  \hline \hline
5  &0.116 & 0.116 & 0.116 &    0.116 \\ \hline
11.46        & 0.114 &0.115  & 0.114 & 0.1155\\ \hline
48.93        & 0.122 & 0.130  & 0.121  & 0.132 \\ \hline
112.14 & 0.139 &0.155 & 0.140 & 0.16\\ \hline
478.74  & 0.238 &0.228 &0.203 & 0.236\\ \hline
1097.3  &0.352 &0.279 &0.250 &0.289  \\ \hline
4684.6 & 0.635 &0.384 &0.338 & 0.395 \\          \hline
10736.8& 0.829 &0.449 &0.390 & 0.461 \\ \hline
20000 &0.985 & 0.499 & 0.429 & 0.512\\ \hline \hline
\end{tabular}
\end{center}
\end{table*}
\end{footnotesize}

G. P.  thanks  the  MIT LNS for  hospitality while this work was being written. R.G. acknowledges support from the Department of Science and Technology, India, under the J.C. Bose fellowship.This work has been partially supported by MEC (FPA2006-05294) and  Junta de Andaluc\'\i a (FQM 101 and FQM 437).
 \begin{footnotesize}
 
 \end{footnotesize}
\end{document}